\documentclass{aa}  
\usepackage{graphicx}
\usepackage{txfonts}
\usepackage{array}
\newcolumntype{C}[1]{>{\centering\let\newline\\\arraybackslash\hspace{0pt}}m{#1}}
\usepackage{hyperref}
\hypersetup{               
    colorlinks=True,
    linkcolor= black,
    citecolor=black
}
\RequirePackage[normalem]{ulem} 
\RequirePackage{color}\definecolor{RED}{rgb}{1,0,0}\definecolor{BLUE}{rgb}{0,0,1} 
\providecommand{\DIFaddbegin}{} 
\providecommand{\DIFaddend}{} 
\providecommand{\DIFdelbegin}{} 
\providecommand{\DIFdelend}{} 
\providecommand{\DIFaddbeginFL}{} 
\providecommand{\DIFaddendFL}{} 
\providecommand{\DIFdelbeginFL}{} 
\providecommand{\DIFdelendFL}{} 
\newcommand{\DIFscaledelfig}{0.5}
\RequirePackage{settobox} 
\RequirePackage{letltxmacro} 
\newsavebox{\DIFdelgraphicsbox} 
\newlength{\DIFdelgraphicswidth} 
\newlength{\DIFdelgraphicsheight} 
\LetLtxMacro{\DIFOincludegraphics}{\includegraphics} 
\newcommand{\DIFaddincludegraphics}[2][]{{\color{blue}\fbox{\DIFOincludegraphics[#1]{#2}}}} 
\newcommand{\DIFdelincludegraphics}[2][]{
        \sbox{\DIFdelgraphicsbox}{\DIFOincludegraphics[#1]{#2}}
        \settoboxwidth{\DIFdelgraphicswidth}{\DIFdelgraphicsbox} 
        \settoboxtotalheight{\DIFdelgraphicsheight}{\DIFdelgraphicsbox} 
        \scalebox{\DIFscaledelfig}{
                \parbox[b]{\DIFdelgraphicswidth}{\usebox{\DIFdelgraphicsbox}\\[-\baselineskip] \rule{\DIFdelgraphicswidth}{0em}}\llap{\resizebox{\DIFdelgraphicswidth}{\DIFdelgraphicsheight}{
                                \setlength{\unitlength}{\DIFdelgraphicswidth}
                                \begin{picture}(1,1)
                                \thicklines\linethickness{2pt} 
                                {\color[rgb]{1,0,0}\put(0,0){\framebox(1,1){}}}
                                {\color[rgb]{1,0,0}\put(0,0){\line( 1,1){1}}}
                                {\color[rgb]{1,0,0}\put(0,1){\line(1,-1){1}}}
                                \end{picture}
                        }\hspace*{3pt}}} 
} 
\LetLtxMacro{\DIFOaddbegin}{\DIFaddbegin} 
\LetLtxMacro{\DIFOaddend}{\DIFaddend} 
\LetLtxMacro{\DIFOdelbegin}{\DIFdelbegin} 
\LetLtxMacro{\DIFOdelend}{\DIFdelend} 
\DeclareRobustCommand{\DIFaddbegin}{\DIFOaddbegin \let\includegraphics\DIFaddincludegraphics} 
\DeclareRobustCommand{\DIFaddend}{\DIFOaddend \let\includegraphics\DIFOincludegraphics} 
\DeclareRobustCommand{\DIFdelbegin}{\DIFOdelbegin \let\includegraphics\DIFdelincludegraphics} 
\DeclareRobustCommand{\DIFdelend}{\DIFOaddend \let\includegraphics\DIFOincludegraphics} 
\LetLtxMacro{\DIFOaddbeginFL}{\DIFaddbeginFL} 
\LetLtxMacro{\DIFOaddendFL}{\DIFaddendFL} 
\LetLtxMacro{\DIFOdelbeginFL}{\DIFdelbeginFL} 
\LetLtxMacro{\DIFOdelendFL}{\DIFdelendFL} 
\DeclareRobustCommand{\DIFaddbeginFL}{\DIFOaddbeginFL \let\includegraphics\DIFaddincludegraphics} 
\DeclareRobustCommand{\DIFaddendFL}{\DIFOaddendFL \let\includegraphics\DIFOincludegraphics} 
\DeclareRobustCommand{\DIFdelbeginFL}{\DIFOdelbeginFL \let\includegraphics\DIFdelincludegraphics} 
\DeclareRobustCommand{\DIFdelendFL}{\DIFOaddendFL \let\includegraphics\DIFOincludegraphics} 

\begin{document} 

   \title{The existence of hot X-ray onsets in solar flares}

   \author{
          Andrea Francesco Battaglia \inst{1, 2}
          \and 
          Hugh Hudson \inst{3, 4}
          \and
          Alexander Warmuth \inst{5}
          \and
          Hannah Collier \inst{1, 2}
          \and
          Natasha L. S. Jeffrey \inst{6}
          \and
          Amir Caspi \inst{7}
          \and
          Ewan C. M. Dickson \inst{8}
          \and
          Jonas Saqri \inst{8}
          \and
          Stefan Purkhart \inst{8}
          \and
          Astrid M. Veronig \inst{8, 9}
          \and
          Louise Harra \inst{2, 10}
          \and
          Säm Krucker \inst{1, 4}
          }

   \institute{
    University of Applied Sciences and Arts Northwestern Switzerland (FHNW), Bahnhofstrasse 6, 5210 Windisch, Switzerland \\
    \email{andrea.battaglia@fhnw.ch}
    \and
    Institute for Particle Physics and Astrophysics (IPA), Swiss Federal Institute of Technology in Zurich (ETHZ), Wolfgang-Pauli-Strasse 27, 8039 Zurich, Switzerland 
    \and
    SUPA School of Physics and Astronomy, University of Glasgow, Glasgow G12 8QQ, UK
    \and
    Space Sciences Laboratory, University of California, 7 Gauss Way, 94720 Berkeley, USA
    \and
    Leibniz-Institut f\"ur Astrophysik Potsdam (AIP), An der Sternwarte 16, D-14482 Potsdam, Germany
    \and
    Department of Mathematics, Physics \& Electrical Engineering, Northumbria University, NE1 8ST, Newcastle upon Tyne, UK
    \and
    Southwest Research Institute, 1050 Walnut Street, Suite 300, 80302 Boulder, USA
    \and
    Institute of Physics, University of Graz, Universit\"atsplatz 5, A-8010 Graz, Austria
    \and
    Kanzelh\"ohe Observatory for Solar and Environmental Research, University of Graz, Kanzelh\"ohe 19, 9521 Treffen, Austria
    \and
    Physikalisch-Meteorologisches Observatorium Davos, World Radiation Center, 7260 Davos Dorf, Switzerland    
             }
             
\authorrunning{A. F. Battaglia et~al.}

   \date{Received March XX, 2021; accepted March YY, 2021}

 
  \abstract
   {
   It is well known among the scientific community that solar flare activity often begins well before the main impulsive energy release. However, a consistent explanation for this phenomenon has not yet been established.
   }
   {
   Our aim is to investigate the earliest phase of four distinct flares observed by Solar Orbiter/STIX and determine the relationships of the newly heated plasma to flare structure and dynamics.
   }
   {
   The analysis focuses on four events that were observed from both Earth and Solar Orbiter, which allows for a comparison of STIX observations with those of GOES/XRS and SDO/AIA. The early phases of the events were studied using STIX and GOES spectroscopic analysis to investigate the evolution of the physical parameters of the plasma, including the isothermal temperature and emission measure. Furthermore, to determine the location of the heated plasma, STIX observations were combined with AIA images.
   }
   {
   The events with clear emission prior to the impulsive phase show elevated temperatures ($> 10\,\mathrm{MK}$) from the very beginning, which indicates that energy release started before any detection by STIX. Although the temperature shows little variation during the initial phase, the emission measure increases by about two orders of magnitude, implying a series of incrementally greater energy releases. 
   The spectral analysis of STIX and GOES from the very first time bins suggests that the emission has a multi-thermal nature, with a hot component of more than $10\,\mathrm{MK}$. Alternative heating mechanisms may be more significant during this phase, since nonthermal emission, as observed by STIX, is only detected later.
   STIX and AIA images reveal the presence of more compact sources of hot plasma early in the flare that originate from different locations with respect to the standard loop-top source that is observed later in the flare. 
   However, because extended bremsstrahlung sources are much more difficult to detect than compact sources, there might be additionally heated plasma in the loop-top during hot onsets. 
   }
   {
   This analysis confirms the existence of "hot onsets," with STIX detecting the hot onset pattern even earlier than GOES. These elevated temperatures imply that energy release actually begins well before any detection by STIX. Therefore, hot onsets may be significant in the initiation, early development, or even prediction of solar flares.
   }


   \maketitle
%

\section{Introduction}

Despite decades of research, the physics behind particle acceleration and plasma heating to millions of degrees in solar flares remains elusive. However, a widely accepted scenario has been developed, the "standard" flare picture, also known as the CSHKP model {(named after \citealt{1964NASSP..50..451C}, \citealt{1966Natur.211..695S}, \citealt{1974SoPh...34..323H}, and \citealt{1976SoPh...50...85K})}.
According to this scenario, there is a sudden release of free magnetic energy in the corona via magnetic reconnection, a large fraction of which is converted into the kinetic energy of high-energy particles. These particles travel along magnetic field lines and release their energy at chromospheric altitudes through heating via Coulomb collisions and radiation through bremsstrahlung emission, which can be observed as X-rays.
The Neupert effect \citep{1968ApJ...153L..59N}, which is consistent with the standard flare model, explains the relationship between soft X-rays (SXRs) and hard X-rays (HXRs). This effect suggests that nonthermal HXRs are emitted by electron beams, while thermal SXRs are derived from the plasma that is heated by the energy deposited by the same electron beams. This plasma subsequently ``evaporates'' and accumulates in the corona.
\citep[e.g.,][]{2005ApJ...621..482V}.

Although observational evidence supports the validity of the Neupert effect, previous statistical studies have found that it may be violated in a significant fraction of events.
When tested in terms of the timing between SXR and HXR emissions, it was found that up to half of the events may show some inconsistencies with this effect \citep[e.g.,][]{1993SoPh..146..177D,1999ApJ...514..472M,2002A&A...392..699V,2005ApJ...621..482V}. 
A rather obvious deviation from the Neupert effect occurs when SXRs are emitted prior to HXRs, which may indicate the presence of thermal plasma not exclusively heated by accelerated electrons.
Although the prior observation of low-energy X-rays may be due to a sensitivity threshold of the HXR detectors \citep{1988SoPh..118...49D}, previous studies \citep[e.g.,][]{1983SoPh...83..267B,2006ApJ...638.1140J} found that this phenomenon cannot be simply explained by a lack of HXR sensitivity \citep{2017LRSP...14....2B}. A statistical study of 503 flares by \citet{2002A&A...392..699V} 
reports thermal ``pre-heating'' events seen in SXRs prior to the onset of the impulsive phase (as detected in HXRs) in more than $90\%$ of the events. Additionally, \citet{2002SoPh..208..297V} found no correlation between the duration of the pre-heating phase and the intensity of the subsequent flare, indicating that pre-heating occurs in the same manner in both weak and intense events. These findings suggest that the Neupert effect may be violated in even more than half of events, at least during the very early phase, regardless of the intensity of the flare.

The onset of flare activity prior to the impulsive phase has been well known for decades \citep[e.g.,][]{1983SoPh...83..267B}. It is important to distinguish between "pre-flare" emission (also known as pre-heating or "flare-preheating") and a "flare precursor."
The term pre-flare commonly refers to the early stages of flares that occur before the start of the main impulsive energy release (i.e., the impulsive phase characterized by a sudden increase in nonthermal emission).
This includes the detection of SXRs before HXRs \citep[e.g.,][]{1983SoPh...83..267B,2002A&A...392..699V,2009A&A...498..891B}. On the other hand, the term flare precursor refers to events that occur well before the flare is observed in X-ray energies, such as nonthermal velocity distributions \citep[e.g.,][]{2001ApJ...549L.245H} or the so-called SXR precursors \citep[e.g.,][]{1991A&AS...87..277T,1996SoPh..168..331F,1998SoPh..183..339F}. We would like to highlight here that in this paper we present an analysis that focuses on the early emission of flares in X-rays (hence, pre-flare). However, as we consider this early emission to be part of the flare itself, we have renamed the pre-flare phase the "flare onset" phase or simply the "onset" phase.

Several studies have investigated this detection of low-energy X-rays prior to HXRs. \citet{1983SoPh...83..267B} conducted a comprehensive study of three events at HXR and radio wavelengths and found radio emission during the onset phase, which indicates the presence of accelerated electrons. 
Additionally, \citet{2009ApJ...705L.143S} found that the energy delivered by nonthermal electrons closely follows (but is not exactly the same as) the 1-8 \AA{} time profile of the X-ray Sensor (XRS) aboard the Geostationary Operational Environmental Satellite (GOES) after conversion to thermal emission.
However, \citet{1992PASJ...44L..71A}, by examining images of ten flares, found evidence for the presence of high-temperature thermal sources in the corona substantially before the impulsive phase. This is in contrast to the assumption of electron-beam-driven evaporation of heated plasma. Similarly, the study by \citet{2009A&A...498..891B} of four distinct events showed that the flare onset phase is characterized by purely thermal emission from a coronal source, with standard nonthermal sources appearing only at the beginning of the main energy release. The authors argue that different heating mechanisms may play a major role during the early stages and that, later on, the resulting heating is transported from the coronal source to the chromosphere via thermal conduction. Coronal emission prior to the impulsive phase has also been reported in other studies \citep[e.g.,][]{2010ApJ...725L.161C,2014ApJ...781...43C,2015ApJ...811L...1C}.
These studies highlight the need for additional heating mechanisms to explain the emission prior to the impulsive phase, beyond the heating provided by accelerated electrons.

\begin{table*}[h]
    \centering
    \caption{Selected events sorted by duration of the onset interval, as measured by STIX, from the very first detection of HXRs to the start of the impulsive phase. All times are given in Earth UT. "SO" in the table stands for Solar Orbiter and "bk" for background.}.
    \vspace{-9pt}
    \fontsize{8.5}{10}\selectfont
    \begin{tabular}{cccccccc}
            \shortstack{ \textbf{Flare}  \\[0pt] \textbf{(Publication)}}  & \shortstack{ \textbf{Onset}  \\[0pt] \textbf{interval}} & \shortstack{ \textbf{Duration}  \\[0pt] \textbf{onset interval}} & \shortstack{ \textbf{GOES}  \\[0pt] \textbf{class}} & \shortstack{ \textbf{GOES bk.}  \\[0pt] \textbf{interval}}& \shortstack{ \textbf{Flare}  \\[0pt] \textbf{location}} & \shortstack{ \\[0pt] \textbf{Distance}  \\[0pt] \textbf{SO - Sun [AU]}} & \shortstack{ \\[0pt] \textbf{Separation angle}  \\[0pt] \textbf{SO - Earth [deg]}} \\ \hline \hline
    \shortstack{ \\[0pt] SOL2021-09-23 \\[0pt] \text{\citep{2023A&A...670A..89S}}} &  \shortstack{ \\[0pt] 15:23:15 \\[0pt] 15:23:28 }      &     $\leq 18\,\mathrm{s}$      &     M1.8  & \shortstack{ \\[0pt] 15:22:30 \\[0pt] 15:23:10 } & E17S36 & 0.60 & 34  \\ \hline
    \shortstack{ \\[0pt]  SOL2021-05-07 \\[0pt] (Mondal et al., in prep.)} &   \shortstack{ \\[0pt] 18:45:55 \\[0pt] 18:48:40 }      &     $165\,\mathrm{s}$      &     M3.9  & \shortstack{ \\[0pt] 18:44:00 \\[0pt] 18:45:40 }  & E76N17   & 0.92 & 97   \\ \hline
    \shortstack{ \\[0pt]  SOL2021-10-09 \\[0pt] \text{\citep{2023arXiv230103650J}}} &   \shortstack{ \\[0pt] 06:24:23 \\[0pt] 06:29:10 }      &     $287\,\mathrm{s}$      &     M1.6  & \shortstack{ \\[0pt] 06:23:00 \\[0pt] 06:24:10 } &  E08N12  & 0.68 & 15  \\ \hline
     \shortstack{ \\[0pt] SOL2022-03-11  \\[0pt] (-)} &  \shortstack{ \\[0pt] 22:14:03 \\[0pt] 22:28:08 }        &     $845\,\mathrm{s}$      &     M2.3 & \shortstack{ \\[0pt] 22:02:00 \\[0pt] 22:02:10 } & W60S24 & 0.45 & 8  \\ \hline
    \end{tabular}
    \label{tab:summary-flares}
\end{table*}

\begin{figure*}[h] 
    \centering
    \includegraphics[width=\textwidth]{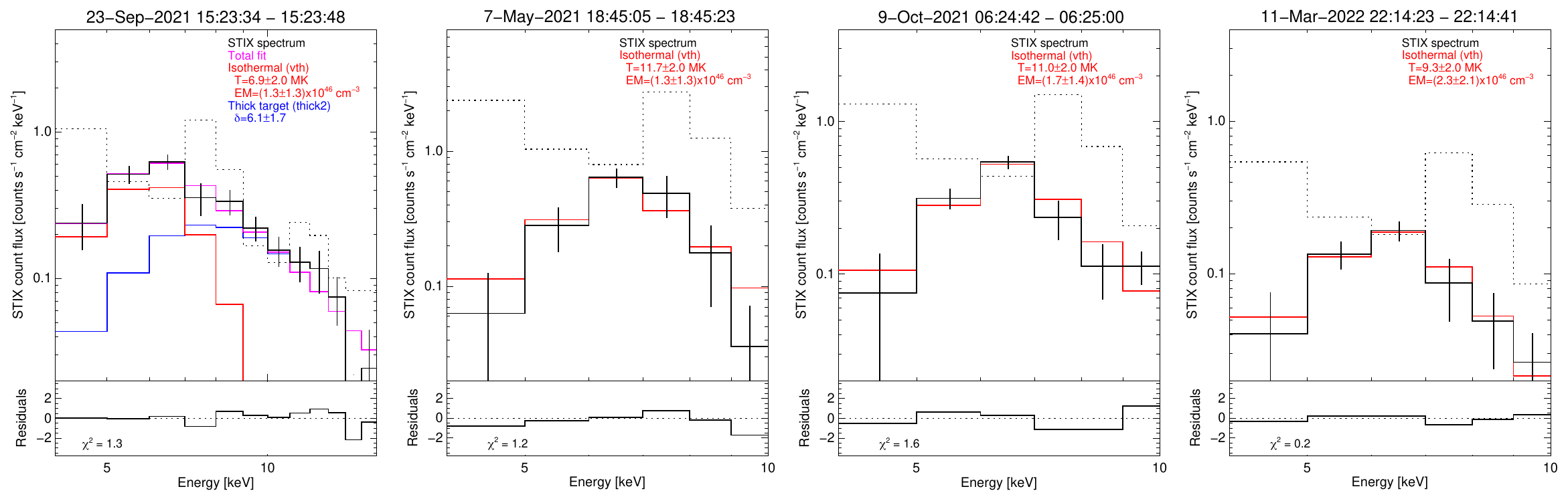}
    \caption{Solar Orbiter/STIX spectral fitting of the earliest spectra for all events under investigation. The solid black line represents the observed background-subtracted flare spectra, and the dashed black lines show the background. The SOL2021-09-23 event (\emph{left}) can be fitted with an isothermal component (red) and an additional nonthermal thick target model (blue). All other events, SOL2021-05-07 (\emph{middle-left}), SOL2021-10-09 (\emph{middle-right}), and SOL2022-03-11 (\emph{right}), can be fitted with a purely isothermal model only. The resulting fitted parameters can be found in the legend.
    }
    \label{fig:spectra-first-bin-all}
\end{figure*} 

A few additional studies that concentrate on the early emission in flares and the location of their source are worth mentioning. \citet{1998ApJ...494L.235A} measured nonthermal velocities, which are particle speeds in the plasma that exceed what would be expected in thermal equilibrium, in a set of events and found the maximum, or the decay, in nonthermal velocity to be prior to the first significant HXR burst. This suggests that the nonthermal velocity may be a direct consequence of the flare energy release process, rather than a byproduct of the energy deposition (like chromospheric evaporation).
\citet{2013ApJ...774..122H} found early nonthermal velocity enhancements located at the base of active regions, which may indicate the activation of flux ropes. 
In the X-ray range, \citet{2001ApJ...560L..87W} reported observations of early brightenings in different locations with respect to the HXR flare footpoints, suggesting that the onset energy release is occurring in loop systems other than those subsequently involved in the flare. 
More recently, in a sample of four events, \citet{2021MNRAS.501.1273H} found significantly elevated temperatures from the very beginning of the flare phase.
This "hot onset" feature appears to be systematic, but a statistical study is necessary to corroborate this.
Based on extreme ultraviolet (EUV) filtergrams from the Atmospheric Imaging Assembly \citep[AIA;][]{2012SoPh..275...17L} on board the Solar Dynamics Observatory \citep[SDO;][]{2012SoPh..275....3P}, they identified this early emission to occur within the footpoints and low-lying loop regions, rather than in coronal structures. 

In this paper we further analyze the geometry of these hot onsets by means of observations performed by the Spectrometer/Telescope for Imaging X-rays \citep[STIX;][]{2020A&A...642A..15K} aboard the Solar Orbiter spacecraft \citep{2020A&A...642A...1M}. STIX's relatively constant background on flaring timescales is advantageous for studies where counting statistics are limited \citep[e.g.,][]{2021A&A...656A...4B,2022A&A...659A..52S,2023A&A...670A..56B}, such as in the very early phase of flares. 
In Sect.~\ref{sec:data-analysis} we describe the event selection and the data analysis. The temporal and imaging observations are presented in Sect.~\ref{sec:results}. In Sect.~\ref{sec:discussion} we discuss the limitations and the interpretation with regard to different scenarios. Finally, we draw our conclusions in Sect.~\ref{sec:conclusions}.


\section{Data analysis \label{sec:data-analysis}}

    
    \subsection{Event selection \label{subsec:event-selection}}
    
    In the present study, we selected four on-disk flare events jointly observed by STIX and GOES, for which the GOES class is larger than M1, in order to ensure good enough counting statistics for both instruments. 
    According to the study of 503 flares by \cite{2002SoPh..208..297V}, on average, the low-energy X-ray emission starts about 3 minutes prior to the impulsive phase. Therefore, we selected two events with this duration shorter than 3 minutes -- SOL2021-05-07T19:00 and SOL2021-09-23T15:30 -- and two with a longer duration of the flare onset interval -- SOL2021-10-09T06:40 and SOL2022-03-11T22:30. Table~\ref{tab:summary-flares} reports relevant information about the flares, including the duration of the flare onset phase as measured by the STIX instrument. The onset phase is defined as the time from the first detected emission to the start of the impulsive phase, when emission above 22 keV rises impulsively. We note that SOL2021-09-23T15:30 is a nice example of an early impulsive flare \citep[e.g.,][and references therein]{2006ApJ...645L.157S,2007ApJ...670..862S}, namely, a flare with short duration low-energy X-ray emission prior to the onset of the main energy release.

    In this paper, X-ray spectroscopy plays a central role. Therefore, to obtain reliable spectral parameters at the very beginning of the event, when counting statistics are intrinsically low, an important selection criterion was to find events during which no other active region was flaring simultaneously. This is a stringent criterion that significantly reduces the number of possible candidates. Indeed, periods of particularly high solar activity have been excluded.
    

   \subsection{Solar Orbiter/STIX data \label{subsec:stix-data}}

    To investigate the early phase of the four selected events, we used the STIX \citep{2020A&A...642A..15K} telescope aboard the Solar Orbiter \citep{2020A&A...642A...1M} mission. Because of the different heliocentric distance of the Solar Orbiter spacecraft to the Sun relative to Earth, the photon arrival time is different. Therefore, in this paper, all times measured by STIX have been corrected in order to take into account this difference and they are expressed in Earth UT.

    To produce the STIX time profiles, we utilized pixel data to avoid the digitization steps caused by compression \citep{2020A&A...642A..15K} and to reduce noise, we integrated the data over 8 or 10 seconds. For spectroscopic analysis, we used spectrogram data because they are continuously available over time, including before the onset of the X-ray emission, and are available at the highest possible cadence. This allows for better flexibility in setting integration intervals for spectroscopy.
    It is worth recalling here that STIX has dynamic time binning in order to optimize onboard memory usage.
    
    Spectral fitting of STIX data has been done using the OSPEX SSWIDL package via the STIX software (version 0.4.0, status March 2023). The very first time bins in which enhanced X-ray emission resulted in meaningful spectral parameters\footnote{This means $\chi^2$ < 2 and at least an isothermal parameter (usually the temperature) with the error smaller than the value itself.} have been fitted manually, as shown in Fig.~\ref{fig:spectra-first-bin-all}, whereas the following ones are fitted automatically using the values of the previous time step as starting parameters. The integration time ranges from 12 to 20 s. This range was chosen to ensure sufficient count statistics for each fitted interval. As reported in Sect.~\ref{subsec:event-selection}, we selected events during which only one active region was flaring. Therefore, as a background for the spectral fitting, we only subtracted the STIX observations taken during quiet times closer to the events (see the dotted curves in Fig.~\ref{fig:spectra-first-bin-all}). These quiet time observations are mostly dominated by the onboard calibration source, especially at the energies considered in this paper (4 - 50 keV), which is constant during flaring timescales. This significantly simplifies the spectral fitting during times where the counting statistics is intrinsically low, such as at the very beginning of the flare.

    For spectroscopic fitting of STIX data, we considered two possible models, either a purely isothermal fit or an isothermal fit with an additional nonthermal component from the standard thick target model \citep{1973SoPh...28..151B}. 
    The nonthermal interpretation of this additional component is consistent with the start of the main energy release.  
    It is important to note that not adding this second component, when high energy counts are clearly present, would result in an over-estimation of the plasma temperature.
    The choice between purely isothermal or the addition of the nonthermal component is based on the $\chi^2$ analysis and the errors on the fitted parameters of the thick target model: if the errors are larger than the values themselves, then we discard this nonthermal component.


    For STIX image reconstruction, the compressed pixel data need to be used. All images have been produced by means of the CLEAN \citep{1974A&AS...15..417H} algorithm, using natural weighting and a beam width of 14.8 arcsec. The size of the selected beam corresponds to the angular resolution of the sub-collimators associated with grid label 3 \citep{2020A&A...642A..15K}, that is, 14.6 arcsec, since the sub-collimators associated with the finest grids, labeled 1 and 2, have been excluded from the analysis. An exception has been done for the thermal loop-top source of the SOL2021-10-09 event. Indeed, at the nonthermal peak, there is no clear modulation in sub-collimators associated with the grid labeled 3 and therefore the beam has been set to the resolution of the sub-collimators associated with the grid labeled 4 (i.e., 20.9 arcsec). The duration of the integration interval has been adapted to the morphology of the flare by having a reasonable amount of total number of counts, which we set to a minimum of 2000 (background-subtracted) for reconstructed images clearly showing two sources. We estimated the error on the position of the onset sources by means of the visibility forward-fit \citep[FWDFIT;][]{2022A&A...668A.145V} algorithm.




    In order to directly overlay the STIX reconstructed images on the AIA maps, the AIA 1600 \AA{} images have been re-projected to the Solar Orbiter vantage point at the time of the events \citep[more details in][]{2021A&A...656A...4B}. In order to co-align the STIX images with AIA, we first applied the standard aspect correction \citep{2020SoPh..295...90W} for the STIX pointing, which is now (as of March 2023) automatically included in the IDL STIX imaging pipeline. For three out of four events under study, an additional manual shift within STIX pointing uncertainty has been applied to the STIX sources to co-align them with the flare ribbons displayed in the re-projected AIA 1600 \AA{} maps. The SOL2021-09-23 flare is the exception and did not require this adjustment. This shift has been determined by co-aligning the X-ray nonthermal sources (22-50 keV) with the flare ribbons during the nonthermal peak. 
    

    \subsection{GOES/XRS data}

    In order to relate the present study with the recent work by \citet{2021MNRAS.501.1273H}, we included the time profiles as well as the isothermal parameters (i.e., temperature and emission measure) obtained from the GOES/XRS, who measures the full-Sun X-ray flux in the $1-8$ and $0.5-4$ \AA{} bands. The background times selected for the subtraction for all events are reported in Table~\ref{tab:summary-flares}. Since the GOES/XRS data analysis is the same here, we refer the reader to Sect.~2.1 of \citet{2021MNRAS.501.1273H} for more details.
    
    We note that the GOES sensors sample the SXR spectrum differently; essentially they bias the photon selection to lower energies, while the STIX response has the opposite bias (see Appendix~\ref{apx:combined-efficiencies}). 
    Accordingly, in the case of multi-thermal sources, we expect systematic differences in the isothermal fits.
    

\section{Results \label{sec:results}}


    \subsection{Time histories}

\begin{figure*}[!] 
        \centering
        \includegraphics[width=\textwidth]{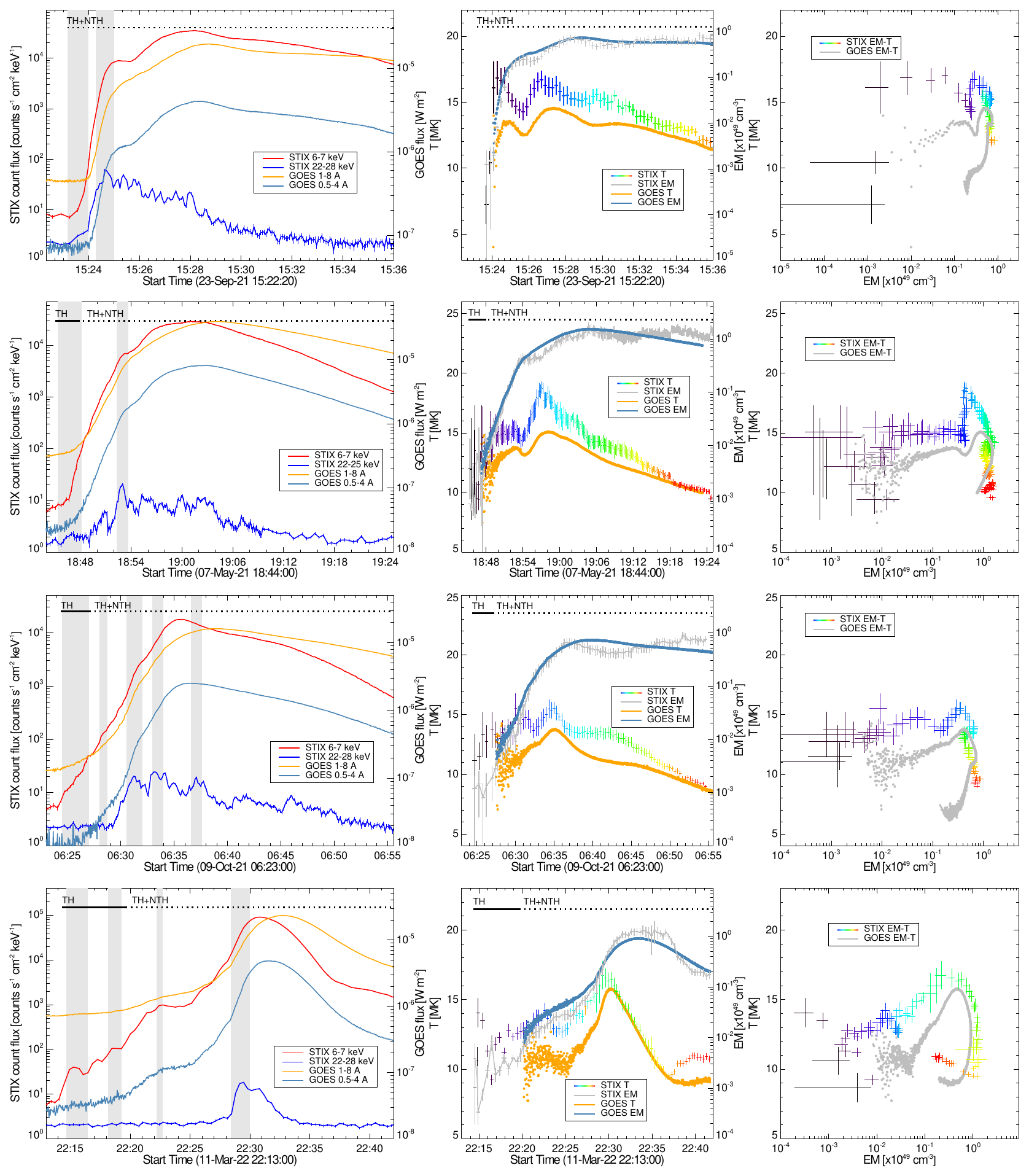}
        \caption{Time histories of the Solar Orbiter/STIX and GOES/XRS fluxes and the isothermal parameters, temperature, and emission measure for all selected events. From top to bottom, the events are sorted by duration of the onset interval: SOL2021-09-23, SOL2021-05-07, SOL2021-10-09, and SOL2022-03-11. From left to right, we show the time evolution of the STIX and GOES fluxes, time profiles of the temperature and  emission measure, and the correlation plot of the temperature as a function of the emission measure. The gray areas in the leftmost panels indicate the accumulation interval selected for producing the STIX images of Figs.~\ref{fig:imaging-Sep23}, \ref{fig:imaging-May7}, \ref{fig:imaging-Oct9}, and \ref{fig:imaging-Mar11}. The horizontal black lines in the plots of the first two columns indicate the spectral model used at different times: purely isothermal (solid, {TH}) or isothermal with an additional nonthermal component (dotted, {TH+NTH}). The STIX emission measure--temperature color coding in the rightmost column matches that of the temperature time series in the central column.
        }
        \label{fig:time-profiles-all}
    \end{figure*}

    The time histories of the four events are shown in Fig.~\ref{fig:time-profiles-all}. This compares STIX isothermal fits with GOES isothermal fits, and in particular plots the emission measure--temperature correlations in the right panels. A comparison of the time histories shows that STIX has greater sensitivity, meaning that its first significant time bin may precede that of GOES.
    
    All events with clear emission prior to the impulsive phase (i.e., SOL2021-05-07, SOL2021-10-09 and SOL2022-03-11) confirm the presence of an early horizontal branch marking the early elevated temperature phase. 
    For these events, the first bins fit well to purely isothermal components, as shown in Fig.~\ref{fig:spectra-first-bin-all}. As indicated by the correlation plots in Fig.~\ref{fig:time-profiles-all}, the temperature has been elevated since the beginning with little variation in the range of 10-16 MK, while the emission measure steadily increases by about two orders of magnitude. A further temperature increase is only observed later during the main energy release.
    A similar behavior is clearly visible in the GOES derived isothermal parameters, with the XRS instrument being more sensitive to lower-temperature plasma than STIX (see Appendix~\ref{apx:combined-efficiencies}). The different temperatures obtained from the two instruments highlight the multi-thermal nature of the emitting plasma.
    We note that the initial elevated temperature is not caused by the instruments reaching their lower limit of temperature sensitivity. In fact, as we can observe later in the flare, the temperature goes even below to what is observed during the flare onset interval for both instruments.

    

    \begin{figure*}[h]
        \centering
        \centering
        \includegraphics[width=0.99\textwidth]{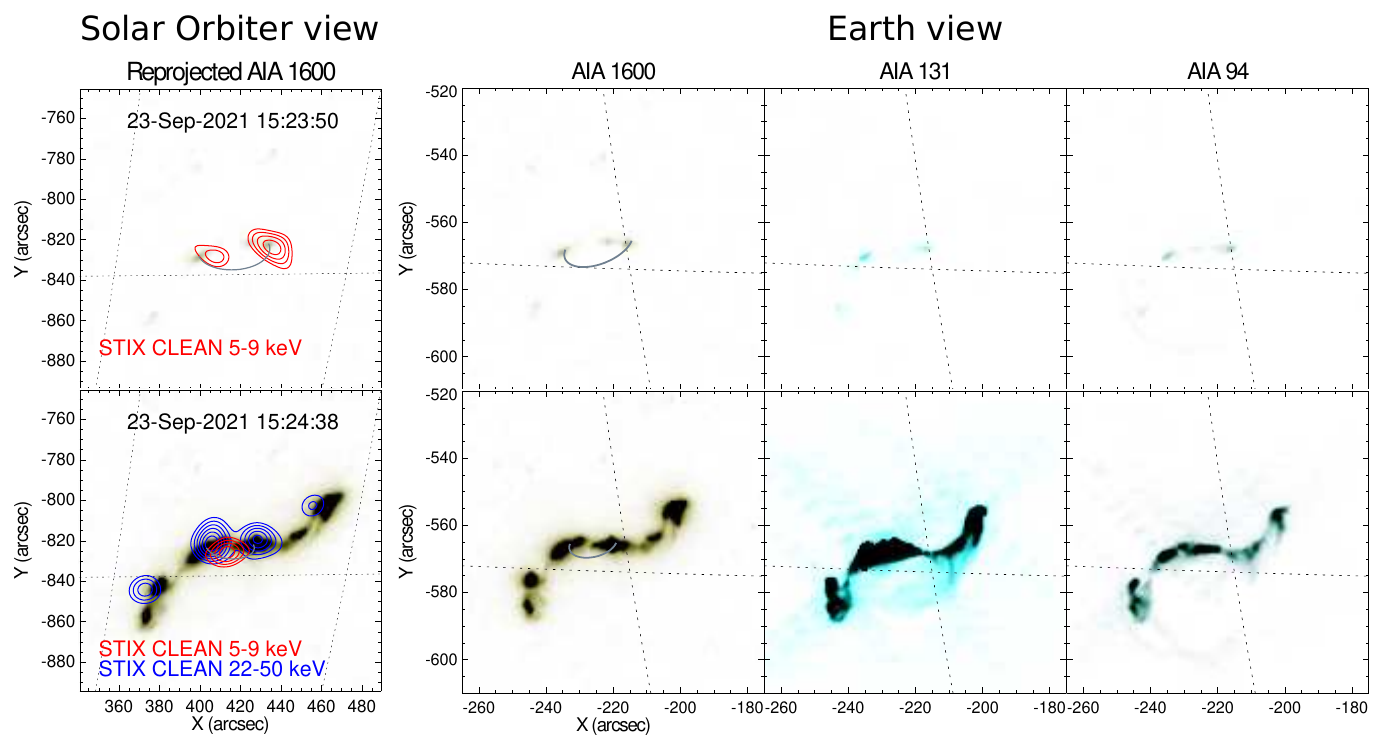}
        \caption{
        Solar Orbiter/STIX and pre-flare-subtracted SDO/AIA images of the SOL2021-09-23 event. The leftmost column shows the STIX reconstructed images, as contour levels, overlaid on the re-projected AIA 1600 \AA{} maps at two different instances: onset (\emph{top panels}) and nonthermal peak (\emph{bottom panels}). The integration intervals used for the STIX images are represented by the gray area in Fig.~\ref{fig:time-profiles-all}. The red contours (60, 70, 80, and 90\% of the maximum) show the images reconstructed within the energy range from 5 to 9 keV, and the blue ones (40, 50, 60, 70, 80, and 90\% of the maximum) from 22 to 50 keV. In order to guide the eye, semicircles perpendicular to the solar surface and connecting the flare ribbons are drawn in gray at the onset and nonthermal peak times. The three columns on the right show the three selected AIA bands, in time (rows). The AIA maps are the closest available to the central time of the STIX integration interval used for reconstructing the images. The same gray semicircles are plotted on top of the AIA 1600 \AA{} maps as seen from Earth.}
        \label{fig:imaging-Sep23}
    \end{figure*}
    


    For the SOL2021-09-23 event, the earliest spectrum could be fitted with an isothermal component together with a nonthermal one, as shown in Fig.~\ref{fig:spectra-first-bin-all}, even though the emission measure is not well constrained. The argument in favor of this interpretation relies on the fact that if we do not add the nonthermal component, the purely isothermal fit returns a temperature of $\sim 25\,\mathrm{MK}$ and the $\chi^2$ is three times larger. It is hard to believe that from an initial temperature of about $\sim 25\,\mathrm{MK}$ there is, after about 30 s, a sudden drop in the plasma temperature to $\sim 17\,\mathrm{MK}$. The same argument applies with a second isothermal, for which the resulting temperature is $\sim 38\,\mathrm{MK}$. In addition, as it is shown by the STIX images (see Sect. \ref{subsubsec:flares1-2}), at higher energies we see footpoints and not coronal sources as it would be expected from typical super-hot components \citep[e.g.,][]{2010ApJ...725L.161C}. In this event the correlation plot (right panel) indicates an increasing temperature with time, which could be interpreted as the nonthermal emission quickly dominating the heating of the plasma from the very beginning.

    \subsection{X-ray and UV/EUV imaging}


        \begin{figure*}[h]
            \centering
            \includegraphics[width=0.99\textwidth]{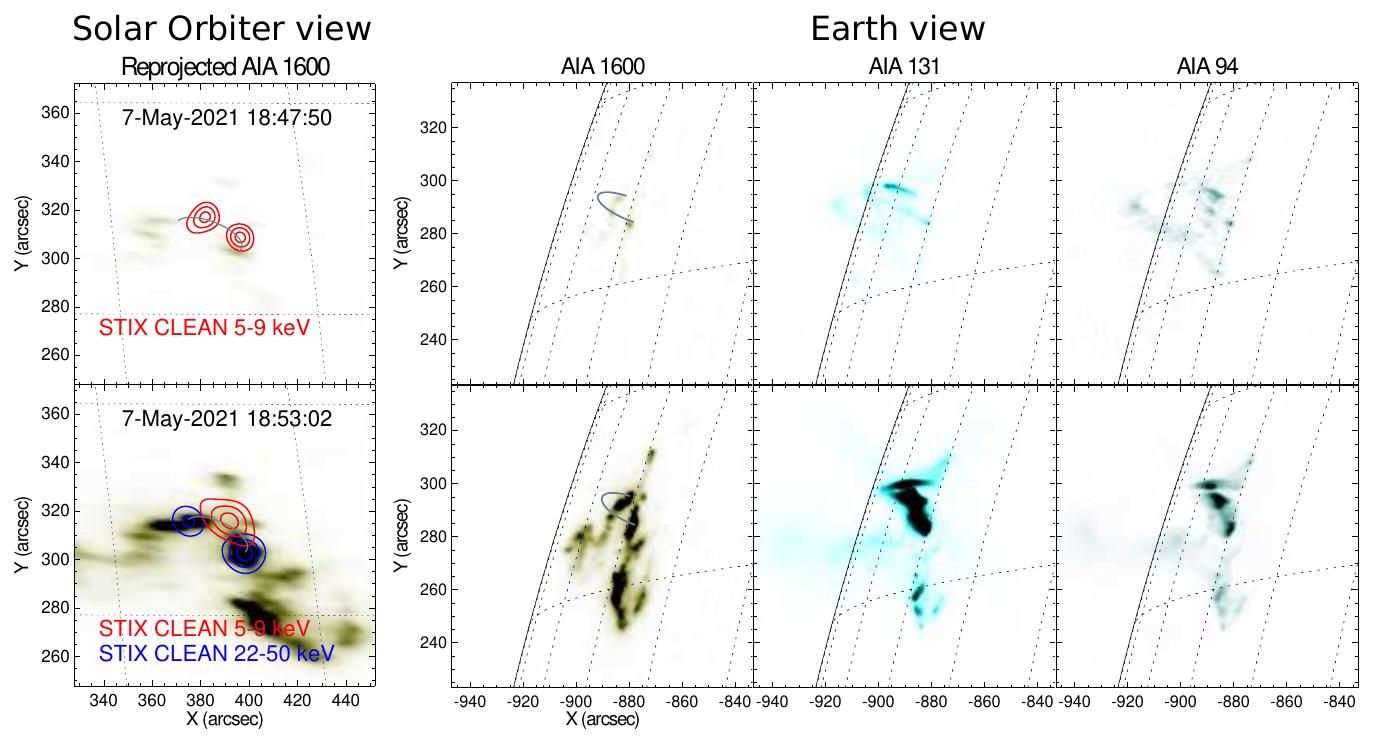}
            \caption{
            Solar Orbiter/STIX and pre-flare-subtracted SDO/AIA images of the SOL2021-05-07 event. The figure has the same format as Fig.~\ref{fig:imaging-Sep23}, except the contour levels of the STIX images are 50, 70, and 90\% of the maximum.}
            \label{fig:imaging-May7}
        \end{figure*}

        \subsubsection{Events with onset intervals shorter than 3 minutes \label{subsubsec:flares1-2}}
        
        Figure~\ref{fig:imaging-Sep23} depicts the AIA and STIX imaging of the SOL2021-09-23 event. This intriguing flare, which has been extensively studied by \citet{2023A&A...670A..89S}, shows at the nonthermal peak (second row) four distinct nonthermal footpoints. \citet{2023A&A...670A..89S} interpret the two inner sources as the flare footpoints, which is indeed consistent with the thermal loop located between them, and the two outer nonthermal footpoints as the anchor points of the eruptive filament. During the onset interval (first row), two distinct sources can be observed in the 5-9 keV STIX image. These sources seem to agree with the corresponding sources visible in the AIA 1600, 131 and 94 \AA{} maps, which indicate the presence of heated plasma at low altitudes. 
        
        The gray loop in the image is a semicircle perpendicular to the solar surface. It has been drawn to guide the eye, as the same loop is plotted in two different projections: as seen from Earth and from Solar Orbiter. The semicircle was obtained by connecting the flare footpoints imaged in X-rays at the nonthermal peak.
        
        The SOL2021-05-07 event displays a similar configuration during the onset interval, as outlined in Fig.~\ref{fig:imaging-May7}. Around the nonthermal peak (second row) the standard picture of solar flares in X-rays clearly stands out, where two nonthermal footpoints correlate well with the chromospheric emission of the AIA 1600~\AA{} maps and the thermal emission with coronal AIA passbands, which outline the top of the flare loop.
        
        In these two flares, the double onset sources have slightly different location with respect to the nonthermal sources, suggesting one of two possible interpretations. On the one hand, their location could come from slightly higher altitudes with respect to the standard nonthermal footpoints, as shown by their relative position with respect to the semicircles perpendicular to the solar surface overlaid on the images. On the other hand, the different position could simply be that the shift is not in altitude, but rather parallel to the solar surface, in which different field lines connects in a different location. 

        It is worth mentioning that the STIX images reconstructed during the onset interval include times where nonthermal emission could be fitted to the spectra. This is obvious for SOL2021-09-23, whereas for the SOL2021-05-07 we could not find a short enough accumulation time for reconstructing a reliable image only including instances with a purely isothermal spectral model. This could be the reason why the STIX images during the onset interval of these two events appear similar (more details in Sect.~\ref{subsec:discussion-limitations}).
    

        \subsubsection{Events with onset intervals longer than 3 minutes \label{subsubsec:flares3-4}}

\begin{figure*}[!h]
            \centering
            \includegraphics[width=0.813\textwidth]{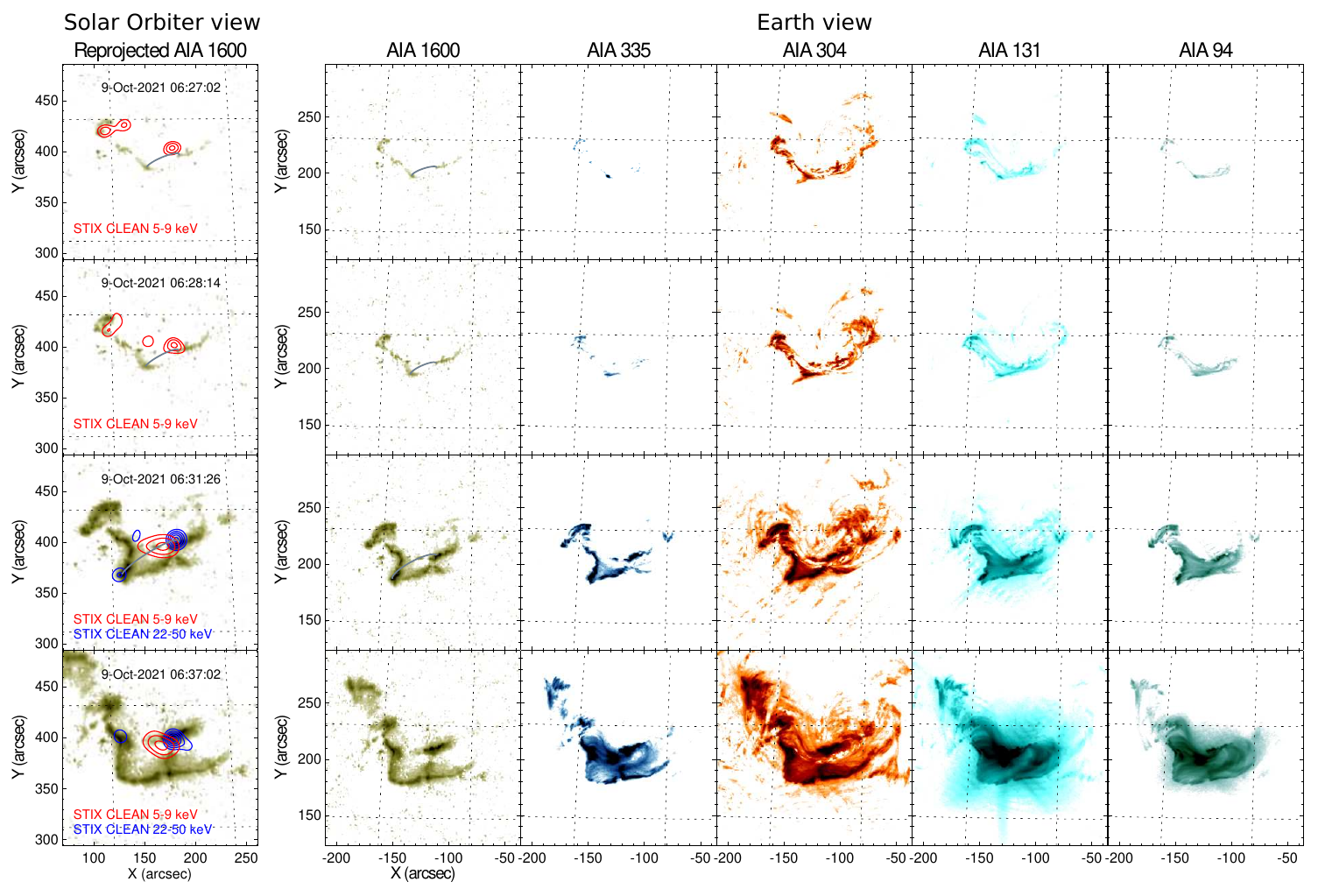}
            \caption{Solar Orbiter/STIX and pre-flare-subtracted SDO/AIA images of the SOL2021-10-09 event. The figure has the same format as Figs.~\ref{fig:imaging-Sep23} and \ref{fig:imaging-May7} except that more time instances and AIA filters have been considered. The contour levels of the STIX images produced in the energy interval from 5 to 9 keV correspond to the 50, 70, and 90\% of the maximum, whereas the ones produced within the 22 to 50 keV range are 50, 60, 70, 80, and 90\% of the maximum.}
            \label{fig:imaging-Oct9}
        \end{figure*}

        SOL2021-10-09 is reported in Fig.~\ref{fig:imaging-Oct9}. Since the onset interval of this event is longer than that of the previously reported ones, it is possible to reconstruct more images prior to the onset of the main energy release. 
        We note mainly two sources in these images. Firstly, there is a northern source that seems to be away from the main flare loop system, as shown by the nonthermal footpoints during the main energy release. Interestingly, this source seems to be spatially correlated with the filament that later erupts. Indeed, from the AIA EUV passbands at 06:31:26, at $(x,y) \simeq (-160,230)$, a structure similar to a filament stands out, which 
        later, at 06:37:02, clearly erupts, as evidenced by the plasma visibly being ejected at $(x,y) \simeq (-180,260)$\footnote{The LASCO CME catalog (\url{https://cdaw.gsfc.nasa.gov/CME\_list/}) reports indeed an eruption associated with this event.}. Secondly, the southern onset source is spatially correlated with the main loop system. The shift with respect to the main thermal source may be due to different field lines lines reconnecting in different locations or to projection effects.
    
\begin{figure*}[!h]
            \centering
            \includegraphics[width=0.813\textwidth]{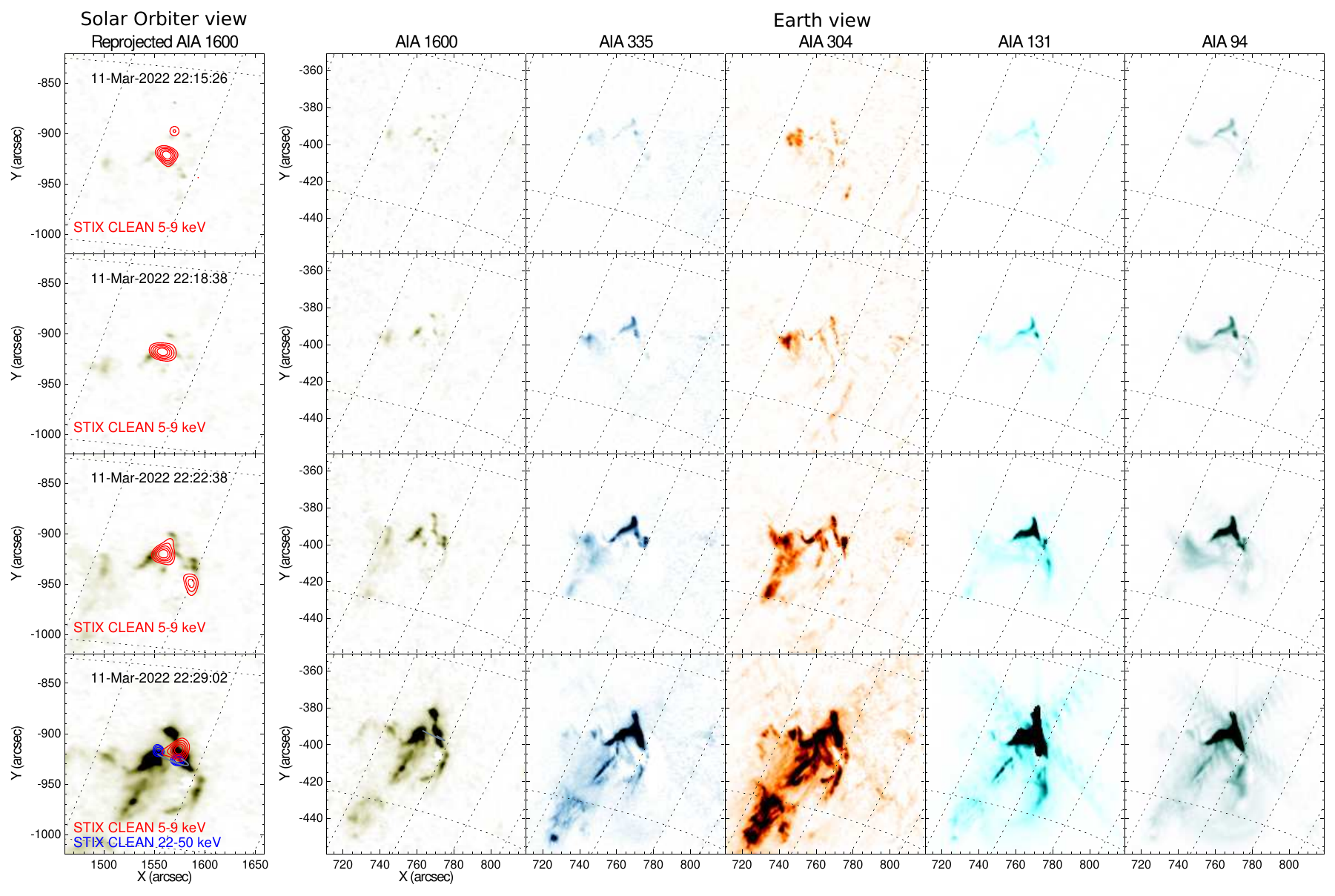}
            \caption{Solar Orbiter/STIX and pre-flare-subtracted SDO/AIA images of the SOL2022-03-11 event. The figure has the same format as Fig.~\ref{fig:imaging-Oct9} except that the contour levels of the STIX image during the onset are 50, 60, 70, 80, and 90\% of the maximum, whereas the ones produced within the 22 to 50 keV range are 60, 70, 80, and 90\% of the maximum.}
            \label{fig:imaging-Mar11}
        \end{figure*}

        Figure~\ref{fig:imaging-Mar11} shows STIX and AIA images of the SOL2022-03-11 event. It is also possible in this case to reconstruct various STIX images during the onset interval. Similarly, different sources are clearly observable during the onset interval with respect to the ones at the nonthermal peak (bottom row). However, since this flare geometry is more compact as compared to SOL2022-10-09, a more detailed analysis on the relative position together with error-bars is needed: We then refer to Sect.~\ref{subsec:locations-sources}. 
        In any case, the double source structure visible in the first STIX image can be related to the small loop structure visible in the AIA 94 \AA{} and (very faint) in the AIA 131 \AA{}. The brighter source, in the center of the figure, persists all along the onset interval. At the time of the third STIX image (third row), another source clearly appears south of the main source. According to the AIA images, this seems to be related to one of the filament footpoints, which later erupts (last row).

    \subsection{Relative location of the X-ray sources \label{subsec:locations-sources}}

\begin{figure*}[!h]
        \centering
        \includegraphics[width=0.99\textwidth]{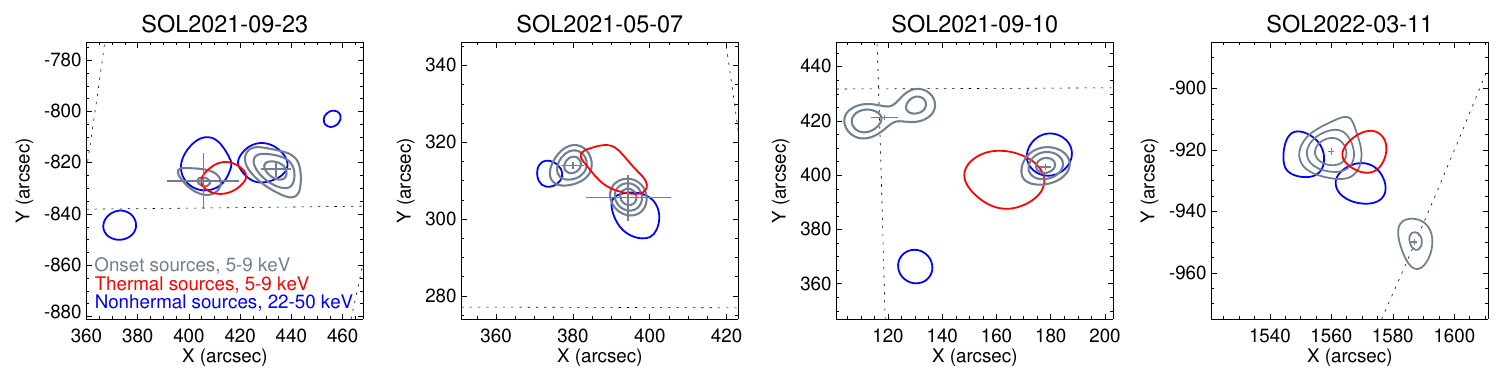}
        \caption{Summary plot with the STIX reconstructed images for all events considered in this paper. Gray contours (50, 70, and 90\% of the peak intensity) represent the images of the sources obtained during the onset interval, and the blue and red contours (50\% of the peak intensity) the nonthermal (22-50 keV) and thermal (5-9 keV) images, respectively, obtained around the nonthermal peaks. The error on the location of the onset sources has been estimated by means of the FWDFIT algorithm, and they are shown as gray crosses.}
        \label{fig:all-STIX-sources}
    \end{figure*}

    In order to properly compare the relative position of the different STIX sources, we plotted them together in the summary plots of Fig.~\ref{fig:all-STIX-sources}.
    The error-bars on the location of the onset sources have been deduced using FWDFIT and then overlaid on the STIX images. 
    
    The common feature in all events analyzed in this paper is that the onset sources are different from the standard single loop-top thermal source that is observed in the same energy range and from the nonthermal peak on. This is clearly shown in Fig.~\ref{fig:all-STIX-sources}. 

\section{Discussion \label{sec:discussion}}


    \subsection{Potential limitations of the analysis} \label{subsec:discussion-limitations}

    In the following, we discuss the limitations for the analysis of the hot onset behavior with STIX and GOES.
    
    Due to telemetry constraints, STIX automatically bins onboard times when flare-related counts are low. This results in time bins ranging from 15 to 20 s during the earliest instances. 
    Therefore, we cannot exclude the possibility that the initial increase in temperature occurs on the order of 15-20 s or less.
    Regardless, we can conclude that the relatively constant temperature behavior observed prior to the main energy release phase is not affected by this dynamic time-binning feature.

    During the STIX observations, the GOES data have 1-second time binning and thus in principle can reduce the STIX limits on the heating timescale of temperature increase. 
    These sensors, unfortunately, have a relatively high and fluctuating background of counts due to radiation-belt electrons \citep[e.g.,][]{2022GOESrept.....M}, and this effectively limits the time resolution on a case-by-case basis.
    As seen in Fig.~\ref{fig:time-profiles-all} from the time-series fluctuations of the GOES-derived isothermal temperature, the STIX data extend the hot-onset pattern to earlier times, and so they provide the best information about the early high temperatures.

    Regarding the spectroscopic analysis, the first detection of nonthermal emission has an inherent uncertainty. In three out of four flares, the low-energy HXR flux was detected before the high-energy HXRs. This could be due to a sensitivity threshold of the HXR detectors \citep{1988SoPh..118...49D} or to the thermal emission completely or partially masking the nonthermal one early in the flare. The fact that the nonthermal component is not well constrained (see Sect.~\ref{subsec:stix-data}), or even not detected, does not necessarily mean that such a component does not exist.

    
    From an imaging perspective, a potential complication arises from the minimum number of counts required to reconstruct a reliable STIX image. This is likely to be a limitation for all events with a relatively short onset interval, as visible from the error-bars in Fig.\ref{fig:all-STIX-sources}, which are larger for these events (i.e., SOL2021-09-23 and SOL2021-05-07) compared to events with a longer onset interval (i.e., SOL2021-10-09 and SOL2022-03-11), where the number of counts included in the reconstructions is significantly higher. Nonetheless, for all STIX images shown in this paper, we included a minimum of 2000 counts (above background) for reconstruction, which should be sufficient to reliably obtain an image with a flare morphology having a maximum of two sources.
    Regarding this aspect, for all events for which we were able to fit a purely isothermal model in the beginning, we attempted to reconstruct at least one image prior to the emission determined to be nonthermal. This worked well for the two events with a longer onset interval but not for the SOL2021-05-07 event, since the integration interval of the first image contains time-bins in which nonthermal emission was clearly observed. Regardless of how we interpret the two sources, this could be the reason why the STIX images during the onset interval of the SOL2021-09-23 and SOL2021-05-07 events appear similar, that is, a double source structure closely associated with the two nonthermal flare-footpoints.

    

    \subsection{Detection of compact versus extended sources in thermal bremsstrahlung \label{subsec:bias-bs}}

    Bremsstrahlung emission has an intrinsic bias that makes it easier to detect compact sources over extended sources. This is because bremsstrahlung is proportional to both the number of particles in the emitting source and the density of the source. For the same number of particles, a compact source has a higher density and therefore appears brighter. To better understand this, consider a plasma with $N$ electrons and $N$ protons at an isothermal temperature $T$. Such a plasma has a thermal energy content of $E_{\mathrm{th}} = 3 k T N$, where $k$ is Boltzmann's constant. The resulting bremsstrahlung flux from this plasma depends on the source volume $V$ and can be written as $F = g(T) \cdot EM$, where $g(T)$ is the function containing the Maxwellian distribution (independent of the volume) and $EM = N^2 / V$ is the emission measure.

    As an example, consider two compact and cylindrical footpoints of 4 arcsec radius and 1 arcsec depth each. The total volume corresponds to $V_{\mathrm{fp}} \simeq 100 \, \mathrm{arcsec}^3$. For a cylindrical loop with a length of 40 arcsec connecting the footpoints, the resulting volume corresponds to $V_{\mathrm{loop}} \simeq 2010 \, \mathrm{arcsec}^3$. For footpoints and a loop source at the same temperature $T$, and where the number of particles summed in the footpoints is the same as that in the loop and corresponds to $N$, the ratio of the resulting bremsstrahlung fluxes is

    \begin{equation}
        \frac{F_{\mathrm{fp}}}{F_{\mathrm{loop}}} = \frac{EM_{\mathrm{fp}}}{EM_{\mathrm{loop}}} = \frac{V_{\mathrm{loop}}}{V_{\mathrm{fp}}} \approx 20.      
    \end{equation}

    \noindent
    In this example, the footpoints have 20 times more flux than the extended loop, despite the coronal loop having the same thermal energy content. The brighter footpoints have a greater impact on images because the lower total flux of the loop is spread over a larger area (i.e., many pixels), while the footpoints' flux is distributed over only a few pixels. To better investigate this, we can consider the flux per pixel, expressed as $F / p$, where $p$ is the number of pixels. The corresponding ratio is

    \begin{equation}
        \frac{F_{\mathrm{fp}} p_{\mathrm{loop}}}{F_{\mathrm{loop}} p_{\mathrm{fp}}} = \frac{V_{\mathrm{loop}} p_{\mathrm{loop}}}{V_{\mathrm{fp}} p_{\mathrm{fp}}}.
    \end{equation}

    \noindent
    Assuming we are looking at the loop from above, with $p_{\mathrm{fp}} = 140$ and $p_{\mathrm{loop}} = 450$, the resulting ratio is $\sim$64, which means that the loop would only be visible at contours below 1.5\% of the footpoint peak flux. Such a large dynamic range in imaging is beyond the range of indirect imaging systems such as STIX; with STIX we generally plot contours down to 10\%, as lower levels reveal the noise of the reconstruction. We note that for a side-on view of the flare, this effect is enhanced. This is because the line-of-sight depth of the loop is the same as that of the footpoints.
    We note that the two equations above are valid only if we assume that $N$ is the same for both the footpoints and loop-top sources, which may not always be true. In such scenarios, the loop may be visible at higher contour levels.
    
    This leads to the question of how this bias influences our observations. For the onset of a flare, compact sources are much easier to detect. Therefore, even if footpoints are detected in STIX imaging, a coronal source hidden in the noise of the image might still have significant energy content. For example, with an imaging dynamic range of approximately 10, a source imaged with less flux than 10\% of the peak flux stays undetected. In other words, if for a certain $F_{\mathrm{loop}}$ we have $F_{\mathrm{fp}}/F_{\mathrm{loop}} > 10$, then the loop is not detected. With the assumed loop geometry, this indicates that the hidden thermal loop could contain up to 50\% more particles than the apparently dominant footpoint sources (i.e., 1.5 times more thermal energy content). Thus, there is a clear bias in detecting footpoints over coronal loops, at least in the very onset of the flare. In the later part of the flare, the flare loops generally dominate. The hot plasma in the heated footpoints expands, and the coronal loop quickly becomes stronger. In combination with the rather low cooling times of coronal loops, this makes the coronal source much brighter than the decreasing footpoint emission later in the flare, despite the intrinsic bias that footpoints are easier to detect.

    The above reasoning for detecting (compact) footpoints, rather than (extended) loops, may apply to events with a relatively short flare onset interval (i.e., SOL2021-09-23 and SOL2021-05-07). In these events, two sources correlated with the nonthermal footpoints are visible at the very beginning, as shown in Fig.~\ref{fig:all-STIX-sources}. For the other two events, a similar argument may apply even in the case of more compact, low-lying loops rather than footpoints.
    
    This bias in thermal bremsstrahlung indicates the need for caution in interpreting X-ray images prior to the onset of the main energy release. 
    To investigate this point further, it would be worthwhile to run simulations in the future and analyze the energy content of the footpoints versus the loop-top.
    Furthermore, combined observations with SXR images, such as those provided by the X-Ray Telescope \citep[XRT;][]{2007SoPh..243...63G} on board Hinode \citep{2007SoPh..243....3K} with a higher imaging dynamic range, should be crucial to find the origin of hot onsets. 

    
   
    \subsection{Interpretation of the emission during the flare onset phase \label{subsec:interpretation-hot-onsets}}

    


        \subsubsection{Early impulsive flare scenario}
    
    The first event in our sample, with the shortest onset interval (SOL2021-09-23), falls in the category of the early impulsive flares.
    
    Its spectroscopic analysis indicates an early increase in temperature during the flare, even though the emission measure is not well constrained. In addition, the first STIX reconstructed image shows a double source structure closely associated with the two nonthermal flare-footpoints, which are the two inner sources.
    One possible interpretation of these sources is the increased plasma densities in the flare loop due to chromospheric evaporation, which causes electrons to be stopped and to emit bremsstrahlung radiation from the legs of the loop \citep{1997ApJ...481..978S,2006ApJ...649.1124L,2006ApJ...645L.157S}. Due to different altitudes with respect to the nonthermal footpoints, they appear to be at slightly different locations due to projection effects. This interpretation is in agreement with the study by \citet{2006ApJ...645L.157S} of the early phases of an impulsive flare, in which low-energy HXR emission was shown to originate from the legs of the flare loop.
    Therefore, in this case, the Neupert effect is consistent from the very beginning. This would explain the increasing temperature observed at the very beginning of the event. 
    Hence, in this flare, the heating caused by energy deposition in the chromosphere due to flare-accelerated electrons may dominate in the early stages of the flare development, in addition to other possible heating mechanisms.
    
    This event nicely showcases the standard flare cartoon for two reasons. First, the geometry of the X-ray and UV/EUV emission closely resembles the standard two-dimensional flare cartoon, as shown in Fig. 9 of \citet{2023A&A...670A..89S}. Second, the energy deposition in the chromosphere by flare-accelerated electrons appears to be the dominant heating mechanism in the early stages of the flare development, which is consistent with the cartoon.
    The evidence for this comes from the absence of significant onset emission, and the quick and impulsive increase in nonthermal emission associated with the main energy release, which potentially overcomes any other possible heating mechanisms that may develop in the very first instances of the flare.

    However, the standard flare picture does not consistently explain the onset emission that is frequently observed in flares. 
    In their analysis of 503 flares, \citet{2002SoPh..208..297V} reported that $90\%$ of the events have a clear low-energy X-ray signal before the onset of the main energy release, indicating that onset emission is common to the vast majority of observations. 
    Hence, we interpret the SOL2021-09-23 event to be a rare and extreme case where electron acceleration works very effectively from the very beginning and may not represent all events that display emission prior to the impulsive phase. In such cases, a process of continued energy release better suits the initial stage as opposed to the impulsive behavior that is typical of the start of the main energy release.
    


        \subsubsection{Flares with clear emission during the onset phase}
        \label{subsubsec:hot-onsets}
        
    The three flares with clear emission prior to the impulsive phase (SOL2021-05-07, SOL2021-10-09, and SOL2022-03-11) show the hot onset behavior.
    
    From the spectroscopic perspective, we note two common features among these events. Firstly, they exhibit a hot temperature from the very beginning. Secondly, this initial elevated temperature shows little variation in the range of 10 to 16 MK throughout the onset interval, while the emission measure steadily increases by up to two orders of magnitude. This clearly indicates that a continuous process of energy input is necessary to explain the observations.
    According to the density-temperature diagram shown originally by \citet{1996SSRv...76..319S} to model a flaring loop (see Fig.~2 in that paper), this phase corresponds to phase b, in which the elevated temperature is maintained by continuous heating. However, the diagram also shows an earlier phase a, a rapid increase in temperature. Now, in accordance with the typical average temperatures of active regions, the question is not whether this phase a is present in flares, because it must be present in order to explain temperatures above $10\,\mathrm{MK}$. Rather, the question is how quickly this phase develops and what densities are involved. Therefore, based on our observations, it appears that neither GOES/XRS nor STIX detect this rapid temperature rise for the three events with clear emission prior to the impulsive phase. This implies that the energy release has already started well before the first signal above background is detected in HXRs and the absence of detected emission is likely due to the low emission measure.
    
    Regarding STIX imaging, the common feature in all events is the presence of onset sources with a different location than the standard loop-top thermal source observed in the same energy range but after the onset of the main energy release \citep[c.f.,][]{2008ASPC..397..130H}. Additionally, for the events with long onset interval, one of the two observed sources is unrelated to the standard flare-loop geometry, and seems to be at, or close to, one of the filament footpoints that eventually erupts.

    As often reported in the literature, the thermal conduction interpretation has been proposed as heating transport mechanism during the early flare phase.
    Previous studies \citep[e.g.,][]{1992PASJ...44L..71A,1993SoPh..146..177D} suggested that some coronal particles may be only slightly energized during the flare energy release, so that this magnetic energy conversion results directly in heating. Consequently, during the onset phase, this coronal heated plasma can ultimately reach the chromosphere by thermal conduction \citep{2017LRSP...14....2B}.
    \citet{2009A&A...498..891B} also reported on events that are consistent with the thermal conduction interpretation. They demonstrated that during the onset phase, the flare morphology displays only one coronal source visible at low X-ray energies, that is, in the thermal range. However, the time range considered by \citet{2009A&A...498..891B} did not included the very early phase as we do in this work and therefore the few minutes difference might have an influence on the fact that only the main loop-top source is detected.
    In our events with long onset intervals, we observe a different situation, with some onset sources clearly located at different positions with respect to the standard coronal thermal source. 
    This suggests that different heating mechanisms and transports may also significantly contribute at these early stages.
 
    

    The outer STIX sources of the two events with longer onset intervals appear to be spatially related to the footpoints of the erupting flux rope.  In this regard, \citet{2001ApJ...560L..87W} found onset sources located in a different location from the flare footpoints as measured in HXRs. Interestingly, \citet{2013ApJ...774..122H} found pre-flare enhancements in nonthermal velocity measurements that were consistent with the location of the footpoints of the flux rope. Additionally, \citet{2018SciA....4.2794J} suggested that turbulence low in the atmosphere could be observed before the heating in the flare is detected. 
    These nonthermal velocity measurements have been observed to peak prior to the first significant burst in HXRs \citep[e.g.,][]{1998ApJ...494L.235A}. This suggests that the nonthermal broadening may be a direct consequence of the flare energy release process, rather than a byproduct of the energy deposition (e.g., chromospheric evaporation). Consequently, it may be that these enhancements in the nonthermal plasma velocities and the hot onsets are related, but we cannot show this here. Further studies are needed to investigate whether the location of the hot onset sources agrees with the pre-flare enhancements of nonthermal plasma velocities. 
    We note that another possible interpretation of the presence of the outer sources may be related to reconnection occurring at a different loop system. However, it is an observational fact that plasma is being ejected from the structures associated with these sources. Future studies on magnetic field extrapolation are needed to answer this question.


        \subsubsection{Heating mechanisms early in the onset phase}

    In the following, we discuss the hot onset behavior in terms of possible different heating mechanisms.
    
    For all events with a clear onset interval, it was possible to fit the first STIX time bins with a purely isothermal model, confirming the presence of heated plasma at temperatures exceeding $10\, \mathrm{MK}$ during the initial phase. Alternative heating mechanisms may be more significant during this phase, as nonthermal emission is only detected later in the onset interval. 
    We should note that the reason why an additional nonthermal component was not always possible to fit may be due to the limited sensitivity of the HXR detectors \citep{1988SoPh..118...49D}. However, as reported by \citet{2017LRSP...14....2B}, this pre-heating problem is already well-known, and previous studies \citep[e.g.,][]{1983SoPh...83..267B,2006ApJ...638.1140J} have found that it cannot simply be explained by a lack of HXR sensitivity. 
    Therefore, it is possible that different mechanisms may play a major role at the beginning by transporting energy from the coronal energy release site to lower altitudes, before electron-beam-driven transport takes over. 
    According to previous studies, other possible mechanisms involved in the early heating could be related to Alfvén waves \citep{2008ApJ...675.1645F}, low-atmosphere turbulence \citep{2018SciA....4.2794J} or Ohmic heating.
    
    Due to the aforementioned reasons, a better understanding of hot onsets is needed, as they may be significant in the initiation and early development, or even the prediction of solar flares.

    


\section{Conclusions and outlook \label{sec:conclusions}}

In this paper we have analyzed the emission during the onset phase of four large flares. These flares were jointly observed by Solar Orbiter/STIX and GOES/XRS, and they presented key aspects that allowed us to draw multiple conclusions. In the following, we summarize our findings.

The three events with clear emission prior to the impulsive phase show the hot onset behavior from the very beginning, as previously reported by \citet{2021MNRAS.501.1273H}. 
The elevated temperature exhibits little variation in the range  10 to 16 MK, whereas the emission measure steadily increases by about two orders of magnitude, which indicates that a process of continued energy release is needed.
Based on typical average temperatures of active regions, there must be a phase before the detection of these hot onsets, on presumably small spatial scales, that explains the increase in temperature to more than $10\,\mathrm{MK}$. The lack of detected emission prior to these enhanced temperatures is likely due to the low emission measure. This suggests that the energy release may have started well before the detection of these hot onsets.

The new results from STIX observations are the following. 
First of all, STIX data extend the hot onset pattern to even earlier times than GOES/XRS, which means it provides better information about the early high temperatures. However, we note that the Reuven Ramaty High-Energy Solar Spectroscopic Imager \citep[RHESSI;][]{2002SoPh..210....3L} is more sensitive than STIX at lower energies (see Appendix~\ref{apx:combined-efficiencies}). Therefore, RHESSI has the potential to detect the actual onset. However, unlike STIX, RHESSI has a strongly varying background that can compromise the analysis during the very early stages. 
Secondly, the spectroscopic analysis, which confirms the presence of heated plasma at temperatures exceeding $10\,\mathrm{MK}$ in the very first stages of the flare, shows that the nonthermal emission is only detected later. This suggests that alternative heating mechanisms besides beam heating may be more significant during this early phase. 
Finally, the combination of EUV/UV images with STIX observations in the very early stages of the flare show the existence of hot plasma at different locations with respect to the standard main thermal loop-top source. However, because extended bremsstrahlung sources are much more difficult to detect than compact sources (see Sect.~\ref{subsec:bias-bs}), there may also be heated plasma at the top of the loop. 

We suggest that future work address two main scientific questions,
firstly, the temperature distribution of these hot onsets. Instruments in the SXR range would be particularly useful in this regard. For example, instruments such as the Solar X-ray Monitor \citep[XSM;][]{2014AdSpR..54.2021V,2020CSci..118...45S} on board the Chandrayaan-2 mission and the Miniature X-ray Solar Spectrometer \citep[MinXSS;][]{2016JSpRo..53..328M,2018SoPh..293...21M} missions, which are sensitive to lower temperatures, could be used to investigate the temperature distribution at lower energies and further explore the initial increase in temperature. Additionally, XRT observations could also be used to diagnose the temperature distribution of hot onset sources at temperatures below the HXR-derived temperatures.
The second would be determining if there is a relationship between nonthermal broadening, which has been observed to precede the low-energy X-ray emission, and these hot onsets. Observations taken by the Interface Region Imaging Spectrograph \citep[IRIS;][]{2014SoPh..289.2733D} and the EUV Imaging Spectrometer \citep[EIS;][]{2007SoPh..243...19C} on board Hinode could be used to investigate whether the location of the hot onset sources agree with the pre-flare enhancements of nonthermal plasma velocities. This would tell us whether turbulent energy contributes to the elevated temperatures observed early in the flare.

\begin{acknowledgements}
      Solar Orbiter is a space mission of international collaboration between ESA and NASA, operated by ESA. The STIX instrument is an international collaboration between Switzerland, Poland, France, Czech Republic, Germany, Austria, Ireland, and Italy.
      
      AFB, HC and SK are supported by the Swiss National Science Foundation Grant 200021L\_189180 for STIX. ECMD, SP and AMV acknowledge the Austrian Science Fund (FWF): I4555-N. AC was supported by NASA grants 80NSSC22M0111, NNX17AI71G, and 80NSSC19K0287.

      This work has benefited from discussions with the STIX team at the "First STIX Team Meeting" held in Prague in April 2023. We express our gratitude to the anonymous referee for the helpful and supportive comments.
\end{acknowledgements}

%
%

\bibliographystyle{aa} 
\bibliography{biblio} 


\begin{appendix}


\section{Responses of X-ray instruments to incoming photon fluxes \label{apx:combined-efficiencies}}


    \subsection{Detection probabilities of incoming photons and effective area}

\begin{figure}
        \centering
        \includegraphics[width=0.5\textwidth]{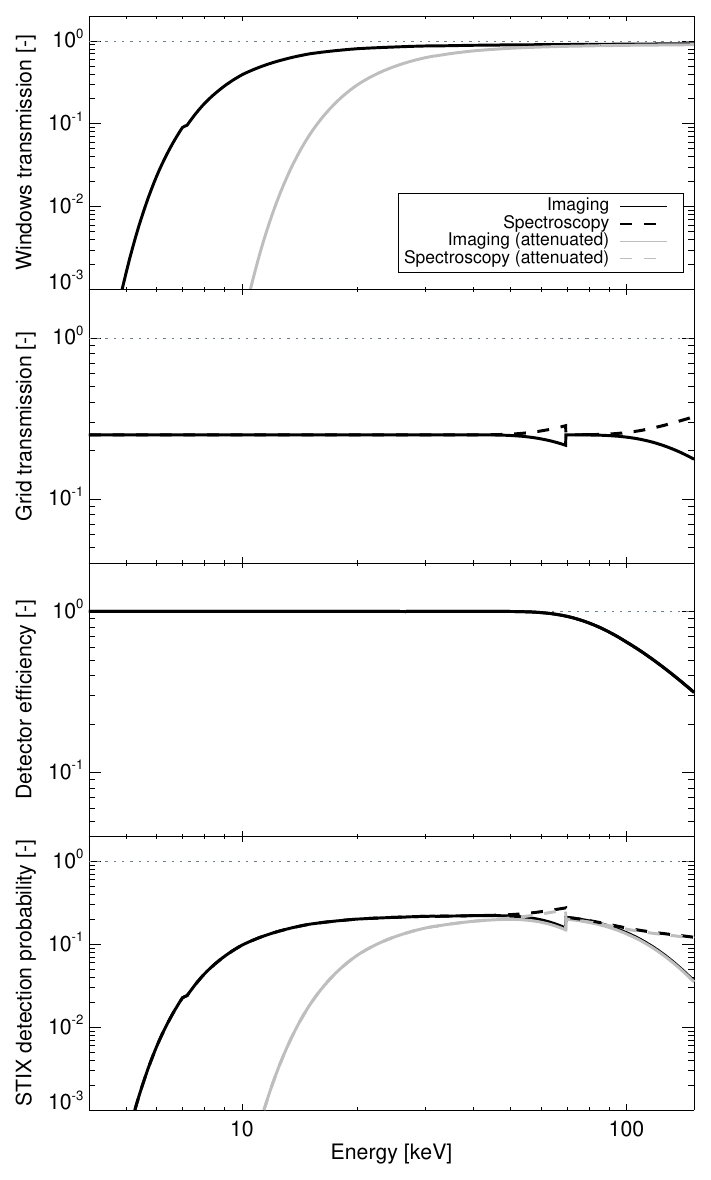}
        \caption{STIX detection probability of an incoming photon flux as a function of energy. From top to bottom, we show the windows' transmission, the grid transmission, the detector efficiency, and the STIX detection probability, which is the convolution of the different components. The solid (dashed) black line shows the STIX imaging (spectroscopy) probabilities. In gray, we highlight the effect of the attenuator insertion. The horizontal dotted line indicates the value of 1.}
        \label{fig:updated-STIX-efficiency}
    \end{figure}

\begin{figure}[!]
        \centering
        \includegraphics[width=0.48\textwidth]{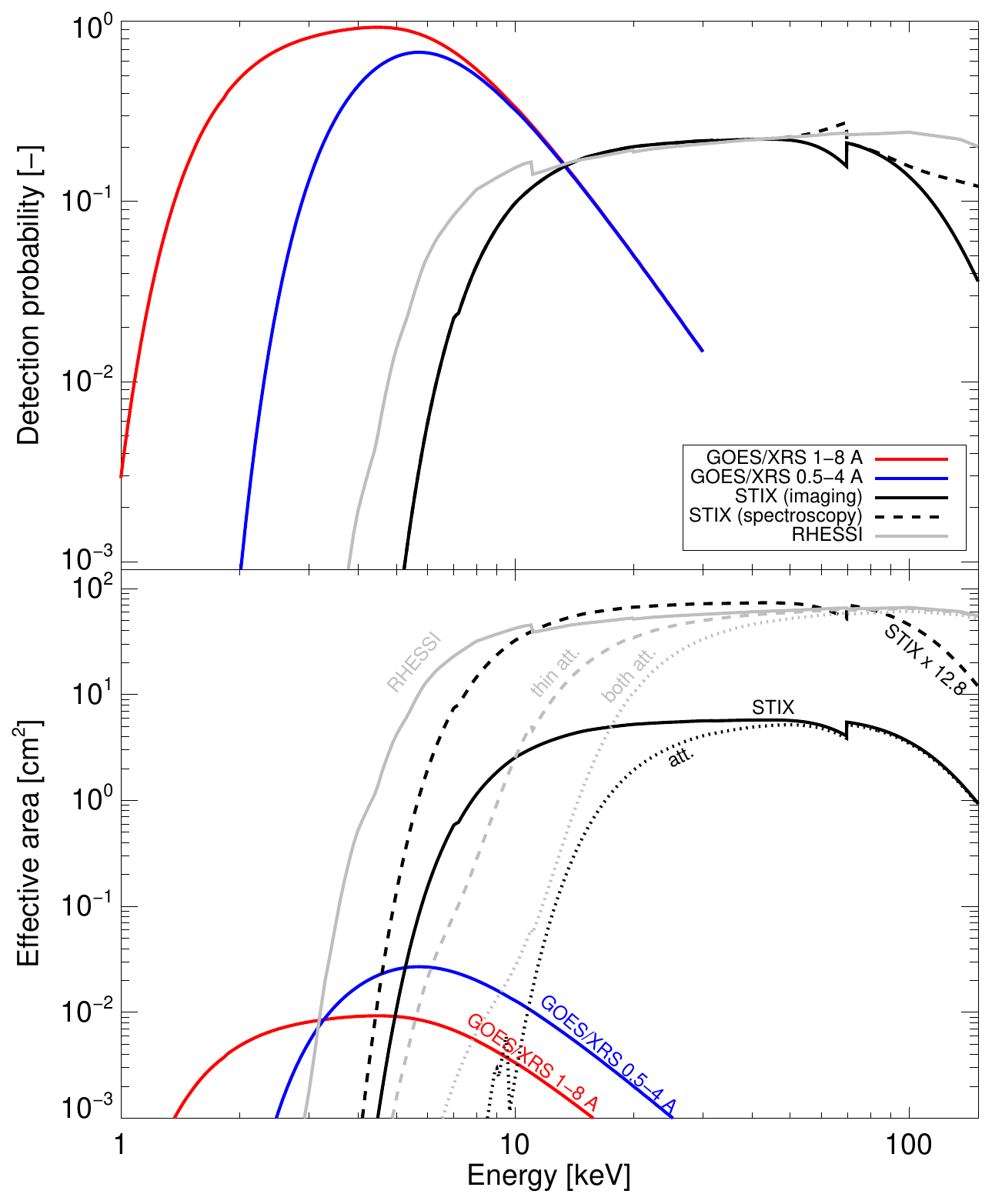}
        \caption{Detection probability of an incoming photon flux and effective area for various X-ray instruments as a function of energy. \emph{Top panel}: Detection probability as a function of energy for GOES $1$-$8$ \AA{} (red), GOES $0.5$-$4$ \AA{} (blue), STIX imaging (solid black), STIX spectroscopy (dashed black), and RHESSI (gray). \emph{Bottom panel}: Effective area as a function of energy for the same instruments. The plot includes the effect of the attenuators in RHESSI (thin attenuator in dashed gray; thin plus thick attenuators in dotted gray) and STIX (dotted black). The dashed black line represents the STIX effective area multiplied by a factor of 12.8 ($ = (0.28 [\mathrm{AU}])^{-2}$), which corresponds to the increased sensitivity at perihelion.}
        \label{fig:stix-goes-responses}
    \end{figure}

    Figure~\ref{fig:updated-STIX-efficiency} displays the updated (as of May 2023) STIX detection probability of an incoming photon flux. This STIX detection probability is the convolution of the transmission through the entrance windows, the tungsten grids efficiency, and the detector efficiency of the STIX CdTe detectors \citep{2020A&A...642A..15K}. 
    In this figure, we highlight the difference in detection probabilities between imaging and spectroscopy, which differ at high energies. This difference is due to the relatively thin grids that become slightly transparent to incoming photons at around 60 keV and above 80 keV. As a result, they do not produce any modulation for imaging but increase the detected photons for spectroscopy. This effect does not occur for RHESSI (see Fig.~\ref{fig:stix-goes-responses}), which has thicker grids \citep{2002SoPh..210....3L}.
    
    The detection probabilities of GOES/XRS and RHESSI (non-attenuated) are shown in the top panel of Fig.~\ref{fig:stix-goes-responses}. For comparison, we report the (non-attenuated) detection probability curves of STIX. The GOES/XRS curves are the result of the combination of the transmission through the entrance windows multiplied by the detector efficiency\footnote{\url{https://data.ngdc.noaa.gov/platforms/solar-space-observing-satellites/goes/goes16/l2/docs/XRS_responsivity/GOES-R_XRS_responsivity_readme.pdf}}. For RHESSI, we considered the effect of the entrance window, the grids and the detector efficiency of the front segment.

    The effective area of GOES/XRS, RHESSI and STIX is shown in the bottom panel of Fig.~\ref{fig:stix-goes-responses}. The difference in the effective area of the GOES/XRS instrument compared to STIX and RHESSI is clearly visible, as they have been designed to work at different X-ray energies. Indeed, flare spectra are much brighter at low X-ray energies (for a tenfold difference in energy, the spectrum is a million times brighter with respect to high X-ray energies), which explains why at low X-ray energies a smaller effective area is enough to detect a sufficient amount of photons. RHESSI has significantly larger effective area at lower energies than STIX and therefore is better suited for detecting hot onsets. Taking observations at perihelion with STIX results in slightly better sensitivity than RHESSI from about 12 keV to 60 keV. Furthermore, the STIX background is stable during flaring timescales \citep{2021A&A...656A...4B,2022A&A...659A..52S,2023A&A...670A..56B} and relatively low between 10 and 20 keV \citep{2020A&A...642A..15K}. These two arguments highlight the increased capabilities of STIX in studying nonthermal emission in microflares during perihelia. 
    
    \subsection{Simulations of flare spectra \label{subsec:simulations}}
    
    \begin{figure}[!]
        \centering
        \includegraphics[width=0.47\textwidth]{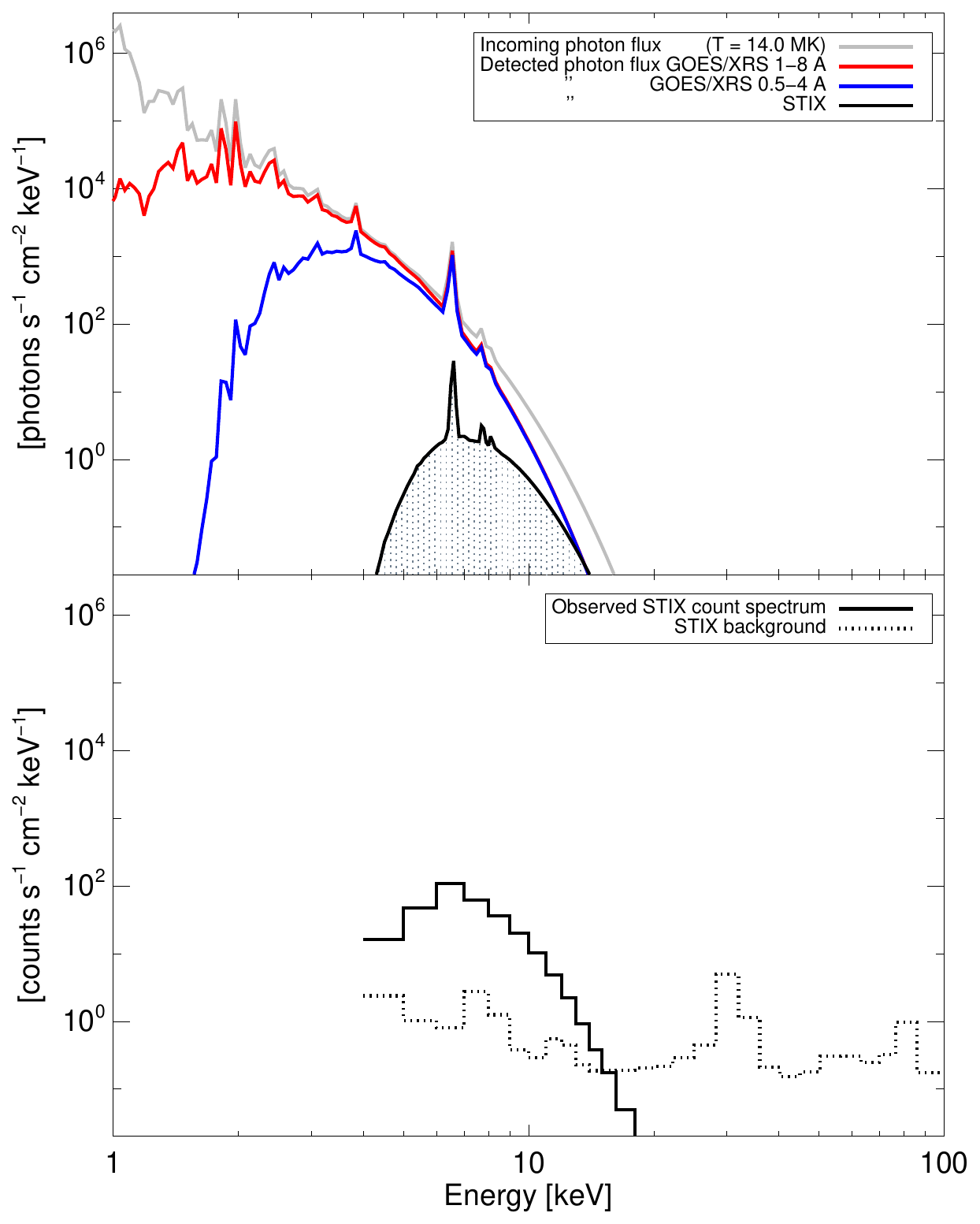}
        \caption{Simulated incoming and detected photon fluxes by STIX and GOES/XRS. \emph{Top panel}: Simulated incoming photon flux, before passing through the entrance windows, for a purely isothermal simulation ($T = 14 \, \mathrm{MK}$ and $EM = 10^{47} \, \mathrm{cm}^{-3}$), shown with the gray curve. Red, blue, and black lines represent, respectively, the corresponding detected photon fluxes of GOES/XRS 1-8 \AA{}, GOES/XRS 0.5-4 \AA,{} and STIX. The hatched area highlights the common area under the different curves, i.e., the photons jointly detected by STIX and GOES. \emph{Bottom panel}: Actual observed STIX count spectrum. In this case, we do not make any distinction between the STIX imaging and spectroscopy spectra since at these energies ($<20\,\mathrm{keV}$) there is no difference.}
        \label{fig:GOES-STIX-simulation}
    \end{figure}

    \begin{figure}[!]
        \centering
        \includegraphics[width=0.47\textwidth]{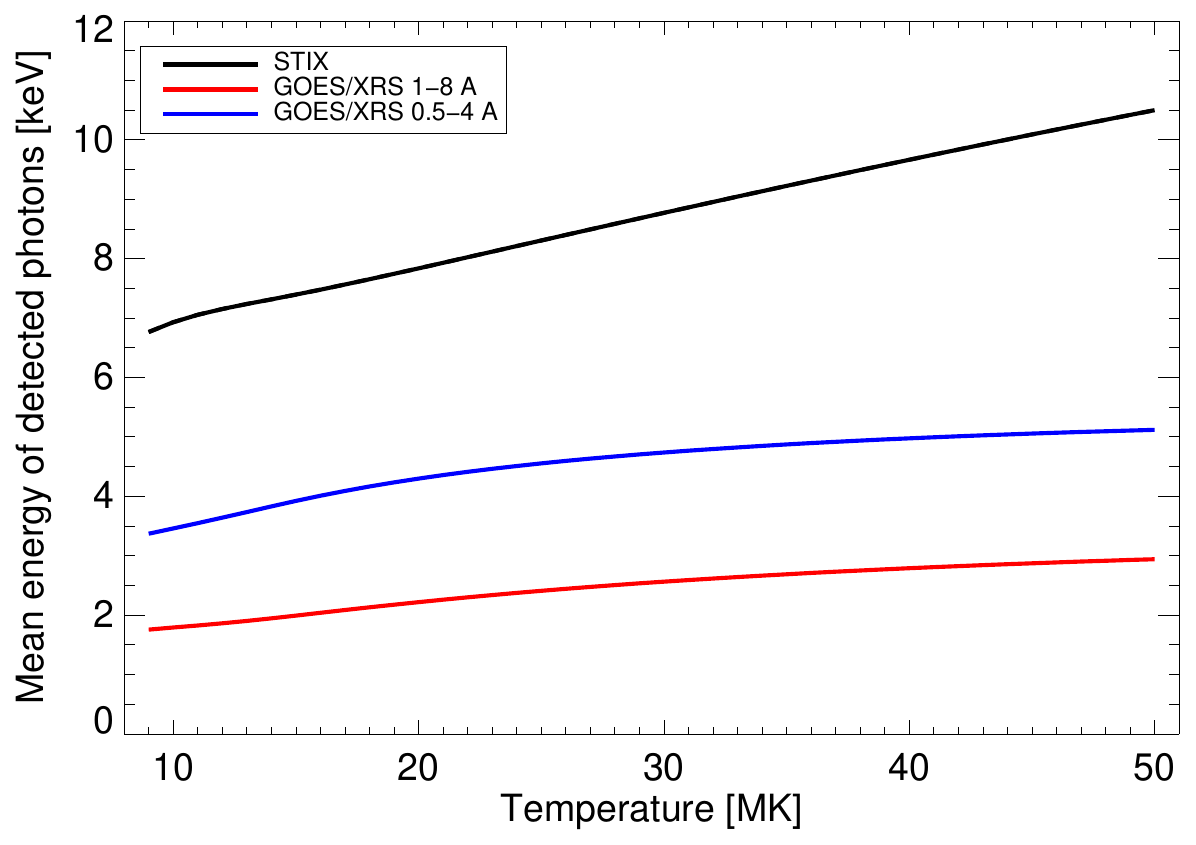}
        \caption{Mean energy of the detected photons of GOES/XRS and STIX against the temperature of the isothermal model. Red and blue curves show the mean energies of the GOES $1$-$8$ \AA{} and $0.5$-$4$ \AA{} bands, respectively, whereas the black line shows the STIX imaging mean energy. For all temperatures, we assumed a constant emission measure of $10^{47} \, \mathrm{cm}^{-3}$. In this case, we do not make any distinction between the STIX imaging and spectroscopy mean energies since, for the considered flare spectra, there is no difference.}
        \label{fig:stix-goes-meanphotonenergy}
    \end{figure}
    
    The top panel of Fig.~\ref{fig:GOES-STIX-simulation} shows the detected photons by STIX and GOES, assuming an isothermal flare spectrum. The detected photon spectra were obtained by convolving the incoming photon flux (gray curve) with the detection probability of the two instruments (see Fig.~\ref{fig:stix-goes-responses}). For the isothermal model, we assumed typical parameters as observed by STIX during the onset interval: $T = 14\,\mathrm{MK}$ and $EM = 10^{47}\,\mathrm{cm}^{-3}$. Additionally, for the STIX detected photon spectrum, we assume that the attenuator is not inserted, as it is the case for all four events analyzed in this paper.

    The bottom panel of Fig.~\ref{fig:GOES-STIX-simulation} displays what STIX measures from the detected photons reported in the top panel. To obtain this curve, the detected photon flux must be convolved with the instrument response matrix. In the case of STIX, the matrix has non-diagonal elements at low energies, which ultimately influences the shape of the spectrum.
    
    The hatched region highlights the common area under the STIX and GOES curves. In practical terms, this means that these are the photons detected jointly by the two instruments, assuming the spectral parameters reported in the first paragraph. From this area, we can deduce that about $99\%$ of the photons detected by STIX are also detected by GOES. However, the percentage of photons detected by GOES/XRS 1-8 \AA{} and 0.5-4 \AA{} that are jointly observed by STIX correspond to only $0.01\%$ and $0.4\%$, respectively.
    This highlights how, for multi-thermal flares, GOES and STIX probe different temperatures.
    
    The difference in the mean energy of the detected photons is illustrated in Fig.~\ref{fig:stix-goes-meanphotonenergy}, in which this quantity is plotted against the temperature of the isothermal model. Assuming isothermal models with temperatures ranging from 9 to 50 MK, the mean energy of the detected photons by GOES/XRS 1-8 \AA{}, GOES/XRS 0.5-4 \AA{}, and STIX is about 2.5 keV, 4.5 keV, and 7-10 keV, respectively.
    This clearly shows that GOES/XRS is more sensitive to lower temperatures compared to STIX.

\end{appendix}


%
\end{document}